\documentclass[a4paper,11pt]{article}
\usepackage{pos}
\usepackage{macros}
\usepackage{slashed}

\title{Unlocking higher-order moments of parton distribution functions from lattice QCD}

\author*[a,b,c]{Andrea Shindler}

\affiliation[a]{%
Institute for Theoretical Particle Physics and Cosmology, 
 TTK, RWTH Aachen University}%
\affiliation[b]{%
Nuclear Science Division, Lawrence Berkeley National Laboratory, Berkeley, CA 94720, USA}%
\affiliation[c]{%
Department of Physics, University of California, Berkeley, CA 94720, USA}%

\emailAdd{shindler@physik.rwth-aachen.de}

\abstract{We present a new method to calculate moments of parton distribution functions of any order with lattice QCD computations. This method leverages the gradient flow for fermion and gauge fields. The flowed matrix elements of twist-2 operators renormalize multiplicatively, and the matching with physical matrix elements is achieved through the use of continuum symmetries. We derive the matching coefficients at one-loop in perturbation theory for moments of any order in the flavor non-singlet case and provide specific examples of operators suitable for lattice QCD computations. The multiplicative renormalization and matching are independent of the choice of Lorentz indices, allowing the use of temporal indices for twist-2 operators of any dimension. This approach should then also significantly enhance the signal-to-noise ratio in the computation of moments.}

\FullConference{31st International Workshop on Deep Inelastic Scattering (DIS2024)\\
 8–12 April 2024\\
Grenoble, France\\}

\begin{document}

\begin{flushright}
    TTK-24-30
\end{flushright}

\maketitle

\section{Parton distribution functions and lattice QCD}
\label{sec:pdf_lqcd}
The connection between parton distribution functions (PDFs) and hadronic matrix elements, which are computable using lattice Quantum Chromodynamics (QCD), is established through the moments of the PDFs, denoted as $\left\langle x^n \right\rangle$. While, in principle, lattice QCD calculations of these moments could allow for a complete reconstruction of the PDFs, this approach remains impractical due to significant theoretical and numerical challenges associated with computing higher moments. Over the past decade and earlier, alternative methods have been proposed to directly determine the full $x$-dependence of the PDFs. For a comprehensive review of these approaches and a complete list of references, see Ref.~\cite{Cichy:2018mum}.
In these proceedings, we summarize a recent proposal~\cite{Shindler:2023xpd} that offers a solution to both the theoretical and numerical challenges that have historically obstructed the direct calculation of moments of any order using lattice QCD.
The standard approach for computing moments of PDFs relies on lattice QCD to determine the hadronic matrix elements of twist-2 operators, defined as
\be 
O_n^{rs}(x) = O^{rs}_{\mu_1 \cdots \mu_n}(x) =
\psibar^r(x) \gmuopen1 \lrDmu2 \cdots \lrDmuclose{n} \psi^s(x)\,,
\ee 
where the $\{ \cdots \}$ indicates symmetrization on the Lorenz indices.
For simplicity, this method is here discussed in the flavor non-singlet case, where $r \neq s$, although extending it to the singlet case poses no particular difficulties. On a hypercubic lattice with spacing $a$, rotational symmetry is reduced to the hypercubic group H($4$), causing the irreducible representations of O($4$) to become reducible under H($4$). This reduced symmetry leads to unwanted mixings during renormalization, as the irreducible representations of H($4$) can mix with lower-dimensional operators and complicate mixing with operators of the same dimension~\cite{Beccarini:1995iv,Gockeler:1996mu}.
For instance, operators with the same number of Lorentz indices, arranged in different combinations, can belong to different irreducible representations of H($4$). A clear example of this is the operator $O_3^{rs}(x)$. If all three Lorentz indices are the same, the operator experiences a power-divergent mixing, proportional to $1/a^2$, with the vector current~\cite{Kronfeld:1984zv}. This issue can be circumvented by choosing an operator with at least two distinct indices~\cite{Martinelli:1987bh}, such as $O^{rs}_{411} - O^{rs}_{433}$. This operator belongs to an irreducible representation of H($4$), and the difference between the two twist-2 operators ensures the subtraction of power divergences. Alternatively, one could select three different indices, for example, $O^{rs}_{124}$. However, in both of these cases, the use of spatial indices necessitates simulations with non-vanishing spatial external momenta when calculating the matrix elements, which worsens the signal-to-noise ratio. The situation becomes even more complex for higher moments, and for $n > 4$, it becomes impossible to avoid mixing with lower-dimensional operators, regardless of the choice of Lorentz indices.

In conclusion, standard methods address the renormalization challenges for $\left\langle x^2 \right\rangle$ and $\left\langle x^3 \right\rangle$ by introducing momentum into the matrix elements, which unfortunately leads to a significant deterioration in the signal-to-noise ratio. The presence of unavoidable power divergences renders these methods ineffective for calculating $\left\langle x^{n-1} \right\rangle$ for $n > 4$.

\section{Flowed fields}
\label{sec:flowmom}

In Ref.~\cite{Shindler:2023xpd}, we introduced a new method that simultaneously addresses both renormalization and signal-to-noise issues in the computation of PDF moments. This method utilizes the gradient flow (GF) for gauge and fermion fields~\cite{Luscher:2010iy,Luscher:2013cpa}, with a review available in Ref.~\cite{Shindler:2022tlx}. 
Gradient flow has become an important tool in lattice QCD computations. One of the key advantages of this approach is that local gauge and fermion fields evolved using GF require only a multiplicative renormalization dependent solely on the fermion content, denoted as $Z_\chi^{1/2}$ for each fermion field. Consequently, any fermion bilinear, irrespective of the gauge content, renormalizes in the same way~\cite{Luscher:2010iy,Luscher:2011bx,Luscher:2013cpa}. This ensures that, at a fixed physical flow-time $t$, the continuum limit for flowed fields is free from additive divergences, assuming proper renormalization of the bare parameters and the flowed fermion fields.

A particularly useful scheme involves the so-called {\it ringed} fermion fields~\cite{Makino:2014taa}, denoted $\rchi$ and $\rchibar$, which are defined by the non-perturbative condition
\be 
\left \langle \rchibar^r(x,t) {\overset\leftrightarrow{\slashed{D}}} \rchi^r(x,t) \right\rangle = - \frac{N_c}{(4 \pi)^2 t^2}\,.
\label{eq:ringed}
\ee 
This ringed scheme is advantageous because it is independent of the regularization method and can be applied both on the lattice and in dimensional regularization. The relationship between the ringed scheme and the $\MSbar$ scheme is known up to next-to-next-to-leading order (NNLO)~\cite{Harlander:2018zpi,Artz:2019bpr}.

\section{Short flow time expansion}
\label{sec:matching}

The new method~\cite{Shindler:2023xpd} is based on the short flow time expansion (SFTX) of flowed fields and on the matching with renormalized fields at vanishing flow time.
At small flow time, $t$, the behavior of the flowed fields, $\rO_i(t)$, in this example defined with ringed fields, is described by an asymptotic expansion~\cite{Luscher:2013vga}
\be 
\rO_i(t)
\underset{\scalebox{0.7}{$t \rightarrow 0$}}{\scalebox{2.0}{$\sim$}}
\sum_i c_{ij}(t,\mu) \left[O_i(t=0,\mu)\right]_{\textrm{R}}\,,
\label{eq:sftx}
\ee 
where the matching coefficients, $c_{ij}(t,\mu)$, are calculated in the same scheme that defines the renormalized fields, $\left[O_i(t=0,\mu)\right]_{\textrm{R}}$, at $t=0$. 
The SFTX in Eq.~\eqref{eq:sftx} thus connects matrix elements of flowed operators with the corresponding renormalized matrix elements at vanishing flow time.
The calculation of the renormalized matrix elements thus typically entails
the lattice QCD computation of the matrix elements with flowed fields. 
Once the flowed fermion fields and the lattice theory are properly renormalized the continuum limit does not present any additional divergence.
The matching coefficients are calculated by inserting the SFTX into off-shell amputated 1PI Green's functions. Technical details on how these type of calculations are performed are given in Refs.~\cite{Mereghetti:2021nkt,Crosas:2023anw}. In the presence of lower dimensional fields contributing to the SFTX the corresponding matching coefficients ought to be calculated non-perturbatively on the lattice. This is the case for CP-odd fields contributing to the neutron electric dipole moment, which are affected by power divergences in the flow time~\cite{Rizik:2020naq,Kim:2021qae,Mereghetti:2021nkt}.

\section{Flowed moments}
\label{sec:flowmom}

The results summarized in the previous section suggest the following strategy.
Let us consider flowed twist-2 fields
\be 
O_n^{rs}(x,t) = \chibar^r(x,t) \gmuopen1 \lrDmu2 \cdots \lrDmuclose{n} \chi^s(x,t)\,,
\label{eq:flowed_t2}
\ee 
where the covariant derivatives are evaluated with flowed gauge fields.
The renormalization of the flowed fields in Eq.~\eqref{eq:flowed_t2} is multiplicative and once defined in terms of ringed fields, have a finite continuum limit. 
It is beneficial now to consider operators belonging to an irreducible representation of O($4$), e.g. traceless and symmetrized rank-$n$ tensors
\be 
\widehat{\rO}_n^{rs}(x,t) = 
\rchibar^r(x,t) \gmuopen1 \lrDmu2 \cdots \lrDmuclose{n} \rchi^s(x,t) - \text{terms~with~}\delta_{\mu_i \mu_j}\,,
\label{eq:flowed_irrep}
\ee
because the matching is multiplicative
\be 
\widehat{\rO}_n^{rs}(t) \underset{\scalebox{0.7}{$t \rightarrow 0$}}{\scalebox{2.0}{$\sim$}} c_n(t,\mu)\left[\widehat{O}_{n}^{rs}(\mu)\right]_{\text{R}}\,.
\ee 

The corresponding matrix element is typically parametrized as 
\be 
\left \langle h(p)| \widehat{\rO}_n(t)
 |h(p) \right\rangle\ = 2
p_{\mu_1} \cdots p_{\mu_n} \left\langle x^{n-1} \right\rangle_h(t)\,,
\label{eq:flowed_me}
\ee 
where for simplicity we omit the flavor indices.
Once the matching coefficients, $c_n(t,\mu)$, are calculated in perturbation theory the moments of the PDFs can be determined inverting the SFTX
\be 
\left\langle x^{n-1} \right\rangle_h(\mu) = \left[c_n(t,\mu)\right]^{-1} \left\langle x^{n-1} \right\rangle_h(t)\,,
\label{eq:flowmom}
\ee 
up to terms with higher powers of the flow time $t$.
There is no restriction on the order of the moments $n$, because $\left\langle x^{n-1} \right\rangle_h(t)$ determined with Eq.~\eqref{eq:flowed_me} have a finite continuum limit for every $n$. 

\section{Matching coefficients}

The matching coefficients are calculable in perturbation theory imposing the matching equations
\be 
\left\langle \psi^r \widehat{\rO}_n^{rs}(t) \psibar^s \right\rangle = c_n(t,\mu)
\left\langle \psi^r \widehat{O}_{n}^{rs,\MS}(t=0,\mu) \psibar^s \right\rangle\,,
\label{eq:matching_eq}
\ee 
where the SFTX is probed with $2$ external unflowed quarks.
The solution~\cite{Shindler:2023xpd} is given by 
\be 
c_n(t,\mu) =  1 + \frac{\gbar^2(\mu)}{\left(4 \pi\right)^2}c_n^{(1)}(t,\mu) + 
O(\gbar^4)\qquad 
c_n^{(1)}(t,\mu) = C_F \left[ \gamma_n \log \left(8 \pi \mu^2 t \right) + B_n\right]
\ee 
where $C_F = 4/3$.
The coefficient of the logarithm is proportional to the 
anomalous dimension of the twist-2 operators~\cite{Gross:1974cs}
\be 
\gamma_n = 1 + 4 \sum_{j=2}^n \frac{1}{j} - \frac{2}{n(n+1)}\,,
\label{eq:an_dim}
\ee 
and provides a check of the calculation.
The finite part 
\bea 
B_n &=& \frac{4}{n(n+1)} + 4 \frac{n-1}{n}\log 2 + \frac{2-4 n^2}{n(n+1)}\gamma_E - \frac{2}{n(n+1)}\psi(n+2) + \\ \nonumber 
&+& \frac{4}{n}\psi(n+1) - 4 \psi(2) - 
4 \sum_{j=2}^n \frac{1}{j(j-1)} \frac{1}{2^j} \phi(1/2,1,j) - \log \left(432\right)\,,
\eea
where $\phi(z,s,a) = \sum_{k=0}^{\infty} \frac{z^k}{(k+a)^s}$, 
provides the new result for the NLO matching.
For $n=2$ we reproduce the result of Ref.~\cite{Makino:2014taa} where the same 
matching has been calculated for the energy-momentum tensor.

\section{Final considerations}
\label{sec:}

Since there are no restrictions on the choice of Lorentz indices, it is convenient to select, for example, in the case of
$n=4$,
\be
\widehat{O}_{4444} = O_{4444} - \frac{3}{4}O_{\left\{\alpha \alpha 4 4 \right\}} 
+ \frac{1}{16} O_{\left\{\alpha \alpha \beta \beta \right\}},
\ee
where repeated indices are summed over.
By subtracting the trace terms, the matching process becomes multiplicative, and the calculation of the hadronic matrix element no longer requires any external spatial momentum. This approach should significantly improve the signal-to-noise ratio, and, combined with the natural smoothing effect of the GF, it offers a solution to the noise issues commonly encountered in standard calculations.

To avoid the direct computation of the ringed fields while still leveraging the correlation among lattice data, a promising strategy is to calculate ratios such as
\be 
\left\langle x^{n-1} \right\rangle^{\MS}_h(\mu) = 
\left\langle x \right\rangle^{\MS}_h(\mu)
\frac{c_{2}(t,\mu)}{c_n(t,\mu)} 
\frac{\left\langle x^{n-1} \right\rangle_{h}(t)}{\left\langle x \right\rangle_{h}(t)}\,, \quad n>2\,,
\ee 
and then relate the results to state-of-the-art calculations of $\left\langle x \right\rangle^{\MS}_h(\mu)$.~\footnote{Results in any other renormalization scheme can be obtained by changing the scheme adopted for the calculation of the matching coefficients.} 

The ratios $\frac{\left\langle x^{n-1} \right\rangle_{h}(t)}{\left\langle x \right\rangle_{h}(t)}$ in the equation above offer several benefits. They reach a finite continuum limit without requiring renormalization of the flowed fermion fields. Additionally, when using Wilson-clover-type fermions, they are $ \mathcal{O}(a)$ improved, aside from short-distance cutoff effects, which are expected to be negligible when calculating the hadronic matrix elements.\footnote{See Ref.~\cite{Kim:2021qae} for an example where these $\mathcal{O}(a)$ short-distance cutoff effects vanish.}
Finally, the ratio of matching coefficients $\frac{c_{2}(t,\mu)}{c_n(t,\mu)}$, which includes next-to-leading log (NLL) resummation, introduces only small perturbative corrections, typically up to around $10\%$~\cite{Shindler:2023xpd}.

In conclusion, we have introduced a new method for calculating moments of any order from lattice QCD. This approach employs an intermediate regulator (the gradient flow), which simplifies the continuum limit. Once O($4$) symmetry is restored, the matching is performed using continuum perturbation theory. The matrix elements of the twist-2 operators can be computed with zero external momenta, and utilizing ratios of matrix elements further improves both the continuum limit and the signal-to-noise ratio.
Initial studies of systematic errors from perturbative matching indicate an uncertainty of approximately $10\%$ or less at NLL, with the potential for systematic reduction by extending the calculation to NNLO. The theoretical limitation on the order of moments that can be calculated in lattice QCD is effectively removed, and we anticipate a significant improvement in statistical uncertainties. Initial numerical results~\cite{Kim:lat24} appear to support this expectation.

\section*{Acknowledgments}
\label{sec:ack}
I acknowledge funding support from Deutsche Forschungsgemeinschaft (DFG, German Research Foundation) through grant 513989149, from the National Science Foundation grant PHY-2209185 and from the DOE Topical Collaboration “Nuclear Theory for New Physics” award No. DE-SC0023663

\bibliographystyle{utphysmod}
\bibliography{refs.bib}

\end{document}